\documentclass[conference]{IEEEtran}
\IEEEoverridecommandlockouts
\usepackage{cite}
\usepackage{amsmath,amssymb,amsfonts}
\usepackage{algorithmic}
\usepackage{graphicx}
\usepackage{textcomp}
\usepackage{xcolor}
\def\BibTeX{{\rm B\kern-.05em{\sc i\kern-.025em b}\kern-.08em
    T\kern-.1667em\lower.7ex\hbox{E}\kern-.125emX}}
\begin{document}

\title{An Item Recommendation Approach by Fusing Images based on Neural Networks}

\author{\IEEEauthorblockN{Weibin Lin}
\IEEEauthorblockA{\textit{College of Computer Science and Technology} \\
\textit{Wuhan University of Technology}\\
Wuhan, China \\
linweibin@whut.edu.cn}
\and
\IEEEauthorblockN{Lin Li}
\IEEEauthorblockA{\textit{College of Computer Science and Technology} \\
\textit{Wuhan University of Technology}\\
Wuhan, China \\
cathylilin@whut.edu.cn}
}

\maketitle

\begin{abstract}
There are rich formats of information in the network, such as rating, text, image, and so on, which represent different aspects of user preferences. In the field of recommendation, how to use those data effectively has become a difficult subject. With the rapid development of neural network, researching on multi-modal method for recommendation has become one of the major directions. In the existing recommender systems, numerical rating, item description and review are main information to be considered by researchers. However, the characteristics of the item may affect the user's preferences, which are rarely used for recommendation models. In this work, we propose a novel model to incorporate visual factors into predictors of people’s preferences, namely MF-VMLP, based on the recent developments of neural collaborative filtering (NCF). Firstly, we get visual presentation via a pre-trained convolutional neural network (CNN) model. To obtain the nonlinearities interaction of latent vectors and visual vectors, we propose to leverage a multi-layer perceptron (MLP) to learn. Moreover, the combination of MF and MLP has achieved collaborative filtering recommendation between users and items. Our experiments conduct Amazon's public dataset for experimental validation and root-mean-square error (RMSE) as evaluation metrics. To some extent, experimental result on a real-world data set demonstrates that our model can boost the recommendation performance.
\end{abstract}


\section{Introduction}
With the massive amount of different source data generated by online services, including e-commerce, social media applications and online news, recommender systems are playing an increasingly important role among them. Matrix factorization (MF) \cite{koren2009matrix,he2016fast}, is one of the most popular collaborative filtering (CF) \cite{su2009survey} techniques, using the numerical rating properly for predicting missing ratings. MF method has achieved great success in the Netflix Prize contest. It casts users and items into a shared latent space, using a latent vector to represent users' interests or items' features. Hence, the result of inner product of these two latent vectors means a user's preferences on a item.

Because of the sparsity of user-item interactions, MF method usually suffers from limited performance. Moreover, without explaining the basic principles, a rating only reflects a user's overall satisfaction towards an item. To improve the accuracy of recommender system, hybrid recommendation \cite{burke2002hybrid,burke2007hybrid} are proposed, which includes two mainly research lines —— the hybridization of algorithms and the hybridization of multi-modal data. For the first research line, researchers attempt to integrate different recommendation methods to improve performance. There are various strategies to combine two methods, such as weighting, switching, mixing, cascading feature combination \cite{burke2007hybrid,gunawardana2009unified} and so on. The combination of content-based \cite{pazzani2007content} and CF-based algorithms is the most common use for research and application. For the other research line, to a certain extent, using more than one information can improve the prediction accuracy of recommender systems, such as rating plus review or rating plus image \cite{he2016ups,he2016vbpr,mcauley2015image}. Researchers propose to integrate numerical rating with textual review for recommendation, including topic modeling \cite{bao2014topicmf,tan2016rating}, and neural network approaches for review modeling \cite{zhang2016collaborative,zheng2017joint}. In order to fuse the characteristics of multi-source data, multi-view learning is a common solution \cite{wang2015robust, wang2018multiview, wang2016iterative}. Although researchers use a variety of data to improve recommendation performance or interpretability, most of them have ignored image information. Currently, the picture is mainly used in the field of retrieval \cite{wang2017effective} and recognition. The appearance of the item is one of the most intuitive and most influential factors of a user's hobby.

Different from the predecessors' works, He et al. \cite{he2016vbpr} integrates image with implicit feedback based on bayesian personalized ranking \cite{rendle2009bpr}. In order to reduce the feature dimension of the picture, authors propose to learn an embedding kernel which linearly transforms the high-dimensional features into a much lower-dimensional space. Compared with the rapid development of neural networks, this method not only has a slower training speed, but also the result is not well. Although Zhang et al. \cite{zhang2016collaborative} used deep neural networks (DNNs) \cite{wu2018deep, wu20193} to model auxiliary information, such as visual content of images. When building models for key collaborative filtering effects, they still use MF method, using an inner product to combine the latent vectors of the users and the items. However, we use neural networks to train model with non-linear method.

In our model, we use explicit feedback as the input, which represents the user's overall satisfaction with the item. The rmse is an evaluation index of our model to predict the missing ratings and train the model. First of all, we extract the visual representations by a pre-trained convolutional neural network (CNN) \cite{wang2015effective} named ImageNet. Secondly, image features are combined with traditional MF method, which can be deemed as a linear transformation model to learn features. However, He et al. \cite{he2017neural} use a MLP model to replace the interaction function and show a well performance gains over traditional methods such as MF. Unlike it's approach, we consider the visual features and plus the features into the model. The last job is to fuse the MF method and MLP, taking into account the image characteristics. The experiments on authoritative dataset like clothing from Amazon.com show that our model outperforms baseline in prediction accuracy. Specifically, our main contributions are listed as follows:
\begin{itemize}
\item We propose a recommended method for fusing images and ratings.
\item We propose a suitable model to incorporate image into predictors of people's preferences based on NCF framework.
\end{itemize}

In the rest of the paper, we first review the related work in Section 2, and present the proposed model in Section 3. Section 4 gives the experimental results and analysis, and Section 5 concludes the paper.

\section{Related Work}
As we all know, collaboration filtering (CF) \cite{su2009survey} algorithm is the most commonly used algorithm in recommendation system. It is divided into three categories, including user-based, item-based and model-based. Among these recommendation algorithms, the model-based recommendation algorithm has become the mainstream of research for a long time. It mainly uses the existing feedback data, such as click, rating, etc, to learn the parameters of model, and then uses this model to predict and recommend. However, MF approach is the most important one in the model-based CF methods. Koren et al. \cite{koren2009matrix} proposes the MF approach based on latent factor model (LFM). The algorithm has an excellent performance in the Netflix competition, which provides a reference for future works.

Because of the sparsity of user-item interactions, CF method usually suffers from limited performance. User-based model is good for recommending popular items in a wide range of interests, but lacks personalization. Item-based model is able to discover long tail items from the users' personal interests, but lacks diversity. From this point of view, each method has its own advantages and disadvantages. In order to combine the benefits of various methods, hybrid model is proposed to make use of different information, including text, context, social \cite{bao2014topicmf,lu2015content,qiao2014combining}, and so on.

The recommended tasks for CF method are mainly divided into two types, one is to predict missing values, and the other is to recommend a short list of items to users named top-k. In the early days, although literature on recommendations primarily focus on explicit feedback \cite{salakhutdinov2007restricted}, recent attention has increasingly turned to implicit data \cite{he2016fast,bayer2017generic}. Depending on the data format used, it can be roughly divided into two ways, which are using implicit feedback to predict the recommended list and using explicit feedback to predict missing scoring values. In our model, we select the latter one.

With the development of neural network, to make use of deep learning, many recent researches have developed non-linear neural network models for CF method \cite{bai2017neural,beutel2018latent,cheng20183ncf,wu2016collaborative,he2017neural}. In particular, instead of using fixed interaction function (i.e., inner product) in MF, He et al. \cite{he2017neural} propose a neural collaborative filtering (NCF) framework to learn the user-item interaction function from data. And then, they propose a model NCF model named NeuMF, which fuse the MF and MLP to learn the interaction function. After that, based on the NCF framework, Bai et al. \cite{bai2017neural} incorporate the neighborhoods of users and items, Cheng et al. \cite{cheng20183ncf} model aspects in textual reviews, and so on.

The appearance of item, such as clothes, is one of the most important factors to influence user's opinions to item. However, in the past, image factors have been ignored because feature extraction methods are failed to achieve effective performance in visual machine learning. However, while deep learning technology has been greatly improved, we can extracted a visual feature from a pre-trained convolutional neural network (CNN) effectively to represent the latent feature of image. Based on the MF method, He et al. \cite{he2016vbpr} propose visually-aware recommender systems. They embedding the high-dimensional image into low-dimensional image by a pre-trained CNN model, which can extract the high performance visual features to use. In the neural network model, they seldom consider the incorporation of rating and image. As a comparison, our focus is to consider visual features based on NCF framework for rating prediction.

\section{METHODOLOGY}
In this section, we introduce three models used in this paper in turn and explain how to generate our models. In order to fuse image and numerical rating for recommendation, we propose the joint model. Table \ref{t1} summarizes the key notations used in the models.
\begin{table}[htbp]
\caption{Notations}
\begin{center}
\begin{tabular}{l l}
\hline
    \textbf{Notation}  &\textbf{Explanation} \\
\hline
    $U,V$       &User set ($|U|$ = $n$), Item set ($|V|$ = $m$) \\
    $R$         &Rating matrix $(n \times m)$ \\
    $y_{ui}$    &Rating assigned by user $u$ to item $i$ \\
    $K$         &Number of latent features in MF \\
    $p_u,q_i,v_i$   &Latent factors of user u, item i, and image i, respectively \\
    $\lambda_u,\lambda_v$    &Regularization parameters  \\
    $\varnothing_{x}$    &x-th neural network layer  \\
    $z_i$    & New item representation  \\

\hline
\end{tabular}
\label{t1}
\end{center}
\end{table}

\subsection{Visual Matrix Factorization (VMF)}\label{sub1}
MF method relates each user and item to a real-valued vector of latent features. Given n users and m products, to decompose rating matrix $R$ $\in$ $\mathbb{R}^{n \times m}$ into two low rank-$K$ matrices $U$ $\in$ $\mathbb{R}^{K \times n}$ and $V$ $\in$ $\mathbb{R}^{K \times m}$, is the goal of MF approach. Let $p_u$ and $q_i$ denote the latent vector for user us and item i, respectively. MF estimates an interaction $y_{ui}$ as the inner product of $p_u$ and $q_i$:

\begin{equation}
\hat{y}_{ui} = {p_u}^Tq_i = \sum_{k = 1} ^K p_{uk} q_{ik}\label {eq1}
\end{equation}

Where $K$ denotes the dimension of the latent space. However, The basic MF approach has a over-fitting problem. The regularized MF method add a normalization factor to the loss function as follows:

\begin{equation}
\min_{U,V} \frac{1}{2}\sum_{u=1}^{n}\sum_{j=1}^{m}(y_{ui} - {p_u}^Tq_i)^2 + \frac{\lambda_u}{2}{||U||_F}^2 + \frac{\lambda_v}{2}{||V||_F}^2 \label{eq2}
\end{equation}

$\lambda_u$ and $\lambda_v$ are regularization parameters for user embedding matrix $U$ and item embedding matrix $V$, respectively, and ${||\cdot||_F}^2$ denotes the Frobenius norm. The gradient descent-based optimization technique is generally applied to find the local minimum solution for Eq. (\ref{eq2}).

Considering that image characteristics can affect user preferences, users who buy clothes will care about its style, color, etc. In order to improve the accuracy of the recommended model, we have added image parameters. The newly loss function as follows:

\begin{equation}
\begin{split}
\min_{U,V} \frac{1}{2}\sum_{u=1}^{n}\sum_{j=1}^{m}(y_{ui} - {p_u}^Tq_i -  {\theta_u}^T\theta_i)^2 \\+ \frac{\lambda_u}{2}{||U||_F}^2 + \frac{\lambda_v}{2}{||V||_F}^2 \label{eq3}
\end{split}
\end{equation}

$\theta_u$ and $\theta_i$ are newly introduced D-dimensional visual factors whose inner product models the visual interaction between $u$ and $i$. To some extent, the user u pays attention to D visual dimensions.  How to use image features effectively in a model has become a research problem. A simple method is to extract features directly from Deep CNN as item features $\theta_i$. However, due to the extracted image features are too high in dimension, the results of the model are unsatisfied. Another approach is to reduce the dimensions of image features. Although some reduction techniques like PCA and CCA are possible to solve this problem, experiments show that these methods lose too much useful information of the original features. Nevertheless, we adopt an embedding kernel \cite{he2016vbpr} to linearly transforms such high-dimensional features into a much lower-dimensional space:

\begin{equation}
\theta_i = Ef_{i} \label{eq4}
\end{equation}

Here E is a D $\times$ F matrix embedding Deep CNN feature space (F-dimensional) into visual space (D-dimensional), where $f_i$ is the original visual feature vector for item $i$.

\begin{figure}[htbp]
\centerline{\includegraphics[height=6cm,width=8cm]{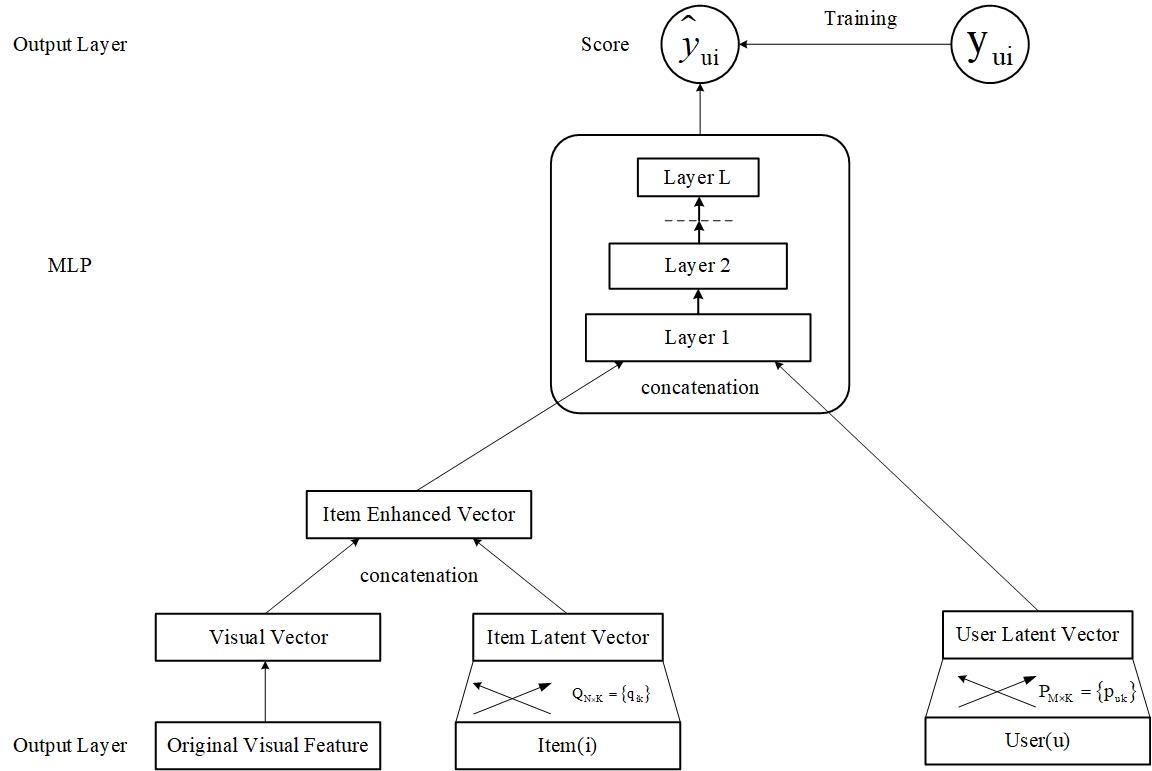}}
\caption{VMLP model}
\label{fig1}
\end{figure}

\subsection{Visual Multi-Layer Perceptron (VMLP)}\label{sub2}
Our model VMLP is based on NCF model, which combine the several pathways to model users and items. As we can see from the first layer of the model in Fig. \ref{fig1}, original visual features has extracted from pre-trained CNN model. At the same time, we use the look-up layer to project the one-hot input of user and item into low-dimensional embedding. The same as VMF, we need to reduce the dimensional of original image features. To address this issue, we propose to add hidden layers on the concatenated vector. In order to integrate the latent vectors of items and image vector, we concatenate these vectors into item enhanced factor. However, to learn and predict users' preferences for items, we use a standard MLP model to train parameters. Unlike the VMF model, we can endow the model a large level of flexibility and nonlinearity to learn the interactions between users and items, rather than a fixed element-wise method. Precisely, the VMLP model is defined as follows

\begin{eqnarray}
\begin{array}{l}
\begin{aligned}
z_1 &= a_1(p_u, q_i) \\
\phi_2(z_1) &= a_2(W_2^Tz_1 + b_2)\\
&...\\
\phi_L(z_{L-1}) &= a_L(W_L^Tz_{L-1} + b_L)\\
\hat{y}_{ui} &= \delta(h^T\phi_L(z_{L-1}))
\end{aligned}
\end{array}
\end{eqnarray}

Where $p_u$, $q_i$, $W_l$, $b_l$, and $a_l$ represent users' factors, items' enhanced factors, the weight matrix, bias vector, and activation function for the $l$-th layer's perceptron, respectively. The latent vector of user and item are fed into a multi-layer neural architecture, which we term as neural network layers, to map the latent vectors to prediction scores. Each layer can be customized to discover certain latent structures of user-item interactions. The dimension of the last hidden layer X determines the model's capability. The final output layer is the predicted score $\hat{y}_{ui}$, and training is performed by minimizing the point-wise loss between $\hat{y}_{ui}$ and its target value $y_{ui}$.

\begin{figure}[htbp]
\centerline{\includegraphics[height=6cm,width=8cm]{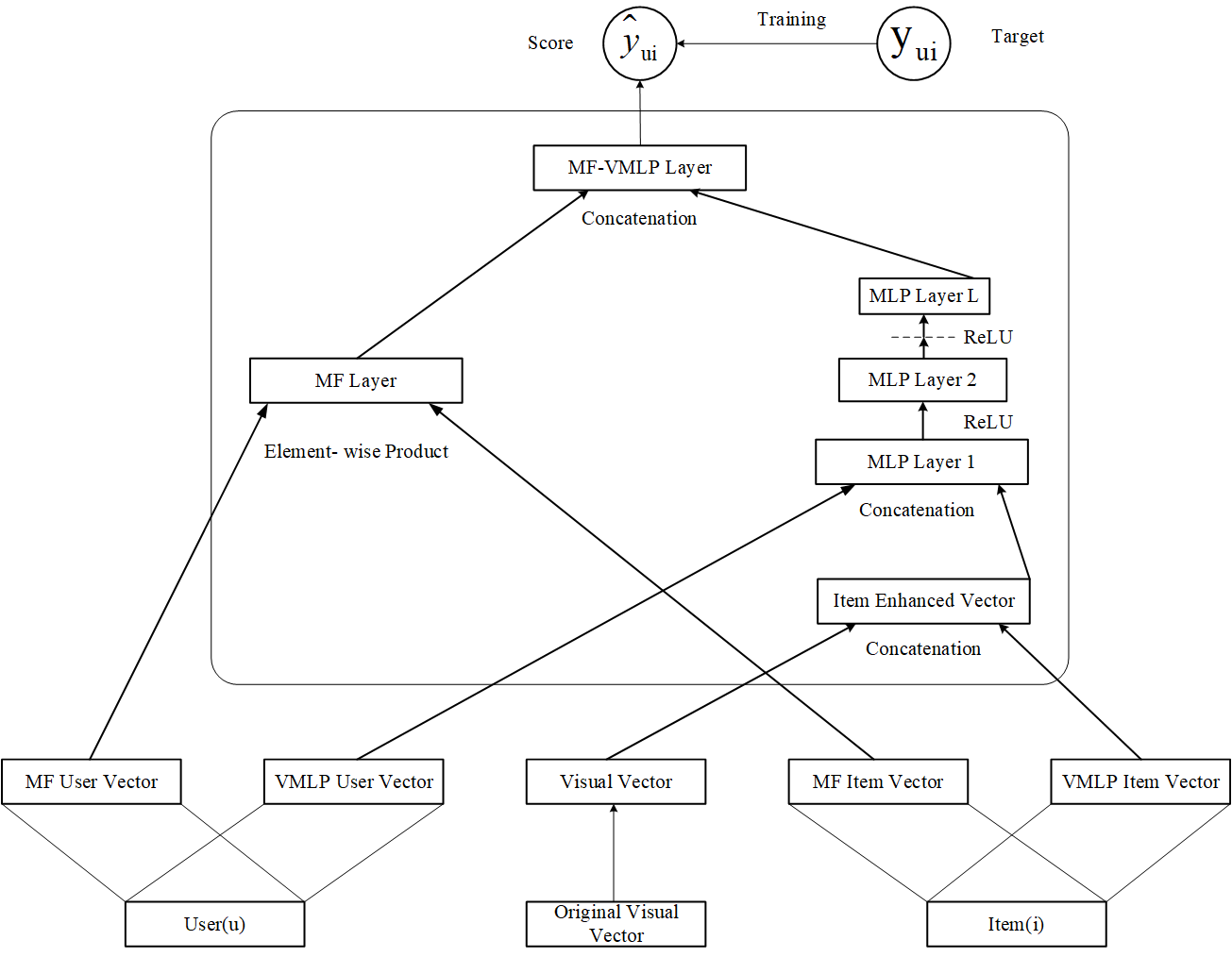}}
\caption{MF-VMLP model}
\label{fig2}
\end{figure}

\subsection{Fusion of MF and VMLP (MF-VMLP)}

Till this moment, we have proposed two model of VMF and VMLP. Through above introduction, VMF uses a linear kernel to model the latent feature interactions, and VMLP applies a nonlinear kernel to learn the interaction function from data. And then, an idea is born that how can we fuse the MF and VMLP based on NCF framework, so as to learn the user-item interactions better. There are two possible ways to solve the issue.

Firstly, one of the easiest ways to work is to share the same input and embedding layers between them, and then combine the outputs of their interaction functions. However, the performance of the fused model might be limited by sharing embedding layers. Once sharing embedding layers, MF and VMLP should use the same size of embedding. While MF and VMLP have their own optimal effect, it's most likely that the embedding layers of both of them are different in size. So this solution is no feasible.

Secondly, to solve the different size from above problem, we let MF and VMLP have their own embedding layers, and concatenate their last hidden layer for combining the two methods. A layer of fully connected layer is added. The model we propose is named MF-VMLP, whose formulations are given as follows

\begin{eqnarray}
\begin{array}{l}
\begin{aligned}
\phi^{MF} &= p_u \odot q_i \\
\phi^{VMLP} &= a_L(W_L^T(a_{L-1}(...a_2(W_2^T\begin{bmatrix}
p_u^V\\
q_i^V
\end{bmatrix} \\&+ b_2)...)) + b_L)\\
\hat{y}_{ui} &= \delta(h^T\begin{bmatrix}
\phi^{MF}\\
\phi^{VMLP}
\end{bmatrix}
\end{aligned}
\end{array}
\end{eqnarray}

Where $p_u$ and $p_u^V$ represent the user embedding for MF and VMLP, respectively, and $q_i$ and $q_i^M$ denote item embeddings. Among several activation functions, we choose the ReLU for MLP layers.

\section{EXPERIMENTS}
In this section, we introduce the dataset and perform the validation results of several models in various datasets. Theoretically, visual appearance is expected to have an influence on users' decision-making process.

\begin{table}[htbp]
\caption{Dataset statistics (after preprocessing)}
\begin{center}
\begin{tabular}{l l l l}
\hline
    Dataset  &\#users  &\#items &\#feedback\\
\hline
    Amazon Women 	&99,748	&331,173		&854,211\\
	Amazon Men		&34,212	&100,654		&260,352\\
	Amazon Phones 	&113,900	&192,085		&964,477\\
\hline
\end{tabular}
\label{t2}
\end{center}
\end{table}

\subsection{Datasets}
The datasets we use are from Amazon.com which introduced by McAuley et al. \cite{mcauley2015image}. We select two categories that are proven to be useful in visual features, named Women's and Men's Clothing. In addition, we also consider Cell Phones, which are expected to paly a little role for models. Each dataset has been process by extracting implict feedback and visual features. We only use the data of user u who has scored more than four items. Table \ref{t2} shows the datasets.

\subsection{Visual Features}
Recall that each item $i$ is associated with a visual feature vector, denoted by $v_i$. To set $v_i$, we use a pre-trained approach to generate visual features from raw product images using the deep learning framework of CAFFE \cite{jia2014caffe}. Following \cite{he2016vbpr}, we adopt the CAFFE reference model with five convolutional layers followed by three fully-connected layers that has been pre-trained on 1.2 million IMAGENET \cite{russakovsky2015imagenet} images. For item $i$, the second fully-connected layer is taken as the visual feature vector $f_i$, which is a feature vector of length 4096.

\subsection{Evaluation Methodology}
We split our dataset into three parts, such as training, validation and test sets. The article uses the root-mean-square error (RMSE) to evaluate the model. The formula is as follows

\begin{equation}
RMSE = \sqrt{\frac{\sum_{u,i \in TestSet}({r}_{u,i} - \hat{r}_{u,i})^2}{TestSet}} \label{eq13}
\end{equation}

Where $r_{ui}$, $\hat r_{ui}$ and TestSet denote real score, prediction score and test set, respectively. As we all know, the lower the rmse, the higher the accuracy of the model.

\begin{table}[htbp]
\caption{RMSE on the test set}
\begin{center}
\begin{tabular}{l l l l l l}
\hline
    Dataset  &MF  &VMF &VMLP &MF-VMLP		&improvement\\
\hline
    Amazon Women 	&1.1303	&1.0886		&1.0639		&1.0575	&7.3\%\\
	Amazon Men		&1.0579	&1.0287		&1.0193		&1.0034	&5.2\%\\
	Amazon Phones 	&1.0883	&1.0794		&1.0854		&1.0879	&0\%\\
\hline
\end{tabular}
\label{t3}
\end{center}
\end{table}

\begin{figure}[htbp]
\centerline{\includegraphics[height=6cm,width=8cm]{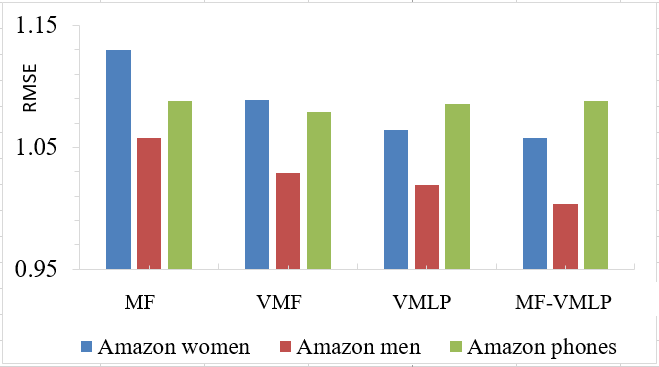}}
\caption{Experimental comparison}
\label{fig3}
\end{figure}

\subsection{Result Analysis}

We can find some details from Table 2. Toward these three datasets, the results show image features affect the users' preferences to some extent. The hybrid recommendation model MF-VMLP gets the best results, it's predicted performance is 7.3\% higher than MF in Amazon Women data. Visual features show greater benefits on clothing than cellphone datasets. We consider that the functional items are not much different in appearance, such as phones, which play a little role on the model. More importantly, neural networks have a large impact on the models.

\subsection{Conclusion And Future Work}
The recommendation system combined with deep learning has become a hot research topic at present. With the rapid development of deep learning, image information becomes more and more important. Visual features influence many of the choices people make. In this paper, we proposed a suitable model to fuse image features and ratings for recommendation. For future research work, due to research on recommendation methods for multi-modal data fusion is also a research hotpot, we will consider more side information to improve the accuracy of recommendation model.


\bibliographystyle{plain}%
\bibliography{bibfile}

\end{document}